# Decarbonizing Basic Chemicals Production in North America, Europe, Middle East, and China: a Scenario Modeling Study

*Tubagus Aryandi Gunawan[1]\*, Hongxi Luo[1], Chris Greig[1], Eric D. Larson[1]*
[1]Andlinger Center for Energy and the Environment, Princeton University, Princeton, New Jersey 08540, United States of America

\*Corresponding author e-mail: arya.gunawan@princeton.edu

**Abstract**

The chemicals industry accounts for about 5% of global greenhouse gas emissions today and is among the most difficult industries to abate. We model decarbonization pathways for the most energy-intensive segment of the industry, the production of basic chemicals: olefins, aromatics, methanol, ammonia, and chlor-alkali. Unlike most prior pathways studies, we apply a scenario-analysis approach that recognizes the central role of corporate investment decision making for capital-intensive industries, under highly uncertain long-term future investment environments. We vary the average pace of decarbonization capital allocation allowed under plausible alternative future world contexts and construct least-cost decarbonization timelines by modeling abatement projects individually across more than 2,600 production facilities located in four major producing regions. The timeline for deeply decarbonizing production varies by chemical and region but depends importantly on the investment environment context. In a best-of-all environments, to deeply decarbonize production, annual average capital spending for abatement for the next two to three decades will need to be greater than (and in addition to) historical "business-as-usual" investments, and cumulative investment in abatement projects would exceed $1 trillion. In futures where key drivers constrain investment appetites, timelines for decarbonizing the industry extend well into the second half of the century.

**Synopsis**: Basic chemicals production has significant carbon emissions that are capital intensive to abate. We model decarbonization pathways considering deep uncertainties under which corporate investment decisions are made.

## 1. Introduction

Chemicals production globally reached nearly $6 trillion of value in 2022 [1] or about 6% of global gross domestic product [2]. Basic chemicals and their derivates are utilized across a vast variety of major industries today [1,3,4], and a review of 15 major studies of the future of the chemicals industry [5] found an average projected growth in chemical industry feedstock demands globally of 2.9% per year to 2050, or roughly a doubling from today. Demand growth is expected to be especially significant in the most rapidly growing economies. The chemical industry accounts for approximately 10% of total global consumption of hydrocarbons today [6], primarily as feedstocks and fuels of fossil origin [3]. The industry as a whole emitted an estimated 2.3 gigatonnes of $CO_2$-equivalent greenhouse gas (GHG) emissions (full value chains, scope 1, 2, and 3) in 2020 [7], or nearly 5% of total global emissions. Chemical producers, especially in economically advanced regions, have shown a strong commitment to decarbonizing the industry [8,9] as have many governments, including the U.S. [10], U.K. [11], and E.U. [12].

To help better understand potential chemical industry decarbonization strategies, this study builds on many prior studies that have investigated chemical-sector emissions abatement technologies – carbon capture and storage (CCS) [13–19], carbon capture and utilization [6,16,20,21], green hydrogen [6,16,17,21], and electrified heating [16,18,20–22] – as well as circular [4,20,21,23–30] or biogenic [15–17,20,21,31] feedstocks, and modeling studies that have explored sector-wide decarbonization pathways. Optimization-based linear programming models that identify least-cost production systems that meet specified exogeneous constraints, especially net-zero emissions by 2050, include the Integrated MARKAL-EFOM System (TIMES) model [32], the Technology Choice Model (TCM) [6,15,20,33], the International Renewable Energy Agency global energy system optimization model [16], and One Earth Climate [34,35], an integrated assessment model. Simulation-based models have also been utilized in major studies, including a carbon intensity model for primary chemicals production developed by the Rocky Mountain Institute [36], an agent-based model [7,37], the HARMONEY model integrating resource consumption with economic dynamics [38,39], and the FORECAST model [17], which primarily focuses on modeling future energy demand, considering factors like policy, technology dynamics, pricing, and other socioeconomic influences.

Here we model decarbonization pathways for production of nine basic chemicals that account for more than 65% of the chemical industry's emissions [6,7,20,22]: ethylene, propylene, benzene, butadiene, toluene, and xylene (collectively, olefins), ammonia, methanol, and chlor-alkali. We include four major producing regions in our analysis that account for an estimated 70% of global production today: North America, Europe, Middle East and China (see SI B1 for the list of countries in each region). We utilize an approach designed to reflect greater real-world pragmatism and fewer abstractions than prior studies to provide new insights. Our work is novel in four key respects.

First, our modeling recognizes the central role that corporate investment decision making will play in decarbonizing any industry in which individual investments are capital-intensive with long development times. Furthermore, the long-lived nature of assets means that investment decisions hinge on their ability to generate durable, long-term returns. Ultimately such returns will depend on the strength of policy and governance, along with consumer demand (and willingness to pay) for sustainable products, both of which are highly uncertain for the decades ahead. Accordingly, we adopt a scenarios approach originally pioneered for corporate strategic planning by Shell [40]. The approach explores how future uncertainties might impact strategic decisions by constructing different narratives around key real-world uncertainties and building quantitative models whose input assumptions are tied to the narratives. In particular we place constraints on the pace of allocation of investment capital for abatement projects to represent different uncertain future investment environments. We note that scenario narratives and associated quantitative modeling outputs are not forecasts of the future but are intended to collectively encompass a range of futures that could eventuate. Second, no prior studies have assessed global decarbonization pathways by aggregating facility-by-facility techno-economic evaluations of potential abatement projects across more than 2600 facilities located in our focus regions. Because every chemical production facility is unique in some respects – capacity, age, input fuels and feedstocks, production process, electricity input carbon-intensity, geographical proximity to abatement resources like suitable $CO_2$ storage capacity, etc. – facility-by-facility analysis is critical for understanding the real-world potential and costs of abatement. Third, we consider future abatement technology cost-learning, as other studies have done, but unlike most prior studies our assumed learning rates quantitatively factor in the feasibility of corporate sharing of experience between abatement projects. Fourth, rather than back-casting from a target of complete decarbonization by 2050 as most prior studies have done, we model the time required to fully decarbonize given various real-world constraints.

Section 2 presents our methodology, including qualitative descriptions of alternative future scenarios and how we quantify these. Section 3 presents and discusses results, including least-cost abatement technology options by specific chemical and region, detailed analysis of steam cracker decarbonization in North America to illustrate the analysis we undertook for each chemical and region, and a global synthesis across all chemicals and regions. Section 4 provides overall conclusions and suggests several directions for future work.

## 2. Methodology

Our scenario narratives qualitatively describe alternative evolutions of sustainability priorities globally over the next half century, and these are assumed to impact the rate of capital invested in abatement, which in turn determines the pace at which decarbonization projects can be deployed toward eventual complete decarbonization of the industry. We allow for a residual level of emissions that are very difficult and costly to abate. For true complete decarbonization, the residuals might be permanently offset by nature-based or engineered $CO_2$ removal measures.

### 2.1. Scenario narratives

Scenario narratives imagine alternative possible future contexts for the global chemical sector's evolution in response to sustainability priorities over the next half-century. To frame the narratives, we selected two key drivers that will heavily influence this evolution, but for which the future direction and strength of change are highly uncertain: 1) strength of governance and coordination on decarbonization and 2) strength of demand for sustainable goods and services. The first factor encompasses many aspects of government involvement and value chain coordination on emissions reduction initiatives. Values at the high end reflect robust emissions reduction and industrial policies plus strong information-sharing among industry players, which helps accelerate cost learning for abatement projects. Low values correspond with limited government involvement and more fragmented approaches to abatement. The second factor reflects consumer demand and willingness to pay for goods and services with sustainable attributes along value chains. At the high end there is strong consumer support for sustainability and high willingness to purchase products with these attributes even at premium prices. Low values reflect incumbent resistance to higher prices and/or mistrust of sustainability claims. Within this two-axis frame, we locate three (non-exhaustive, but broadly representative) scenarios (Figure 1a).

The Sustainable United (SU) scenario is a best-possible sustainability scenario characterized by strong global governance and widespread demand for sustainable products. The deeply collaborative environment sees governments, businesses, and NGOs united behind the climate agenda and broader sustainability goals. Strong government coordination facilitates deep collaboration across regions and economic sectors, with proactive regulators reducing barriers to corporate collaboration. Coalitions comprising government, industry, and environmental/social NGOs design policy, regulatory, and incentive measures that drive sustainable change. Digital innovation facilitates technology exchange that helps rapidly drive down the cost of emission abatement technologies and establish confidence in low carbon product attribution. Consistently strong governance builds institutional trust and underpins acceptance of technological change, leading to high levels of willingness to pay for low-emission energy and products.

In the Green Authority (GA) scenario, much like the SU scenario, strong global governance drives progress through regulation and public-private collaboration, but, unlike SU, consumers and communities remain skeptical and resistant to change. Strong governance still enables substantial progress. The chemical industry sees broad alignment on measurement, tracking, and certification frameworks, and some cost reductions in clean electricity and heat supply systems are achieved, albeit at a slower pace than in the SU scenario. However, societal distrust in governments and corporations manifests in grassroots opposition and legal challenges that, inter alia, slow the deployment of emission abatement projects.

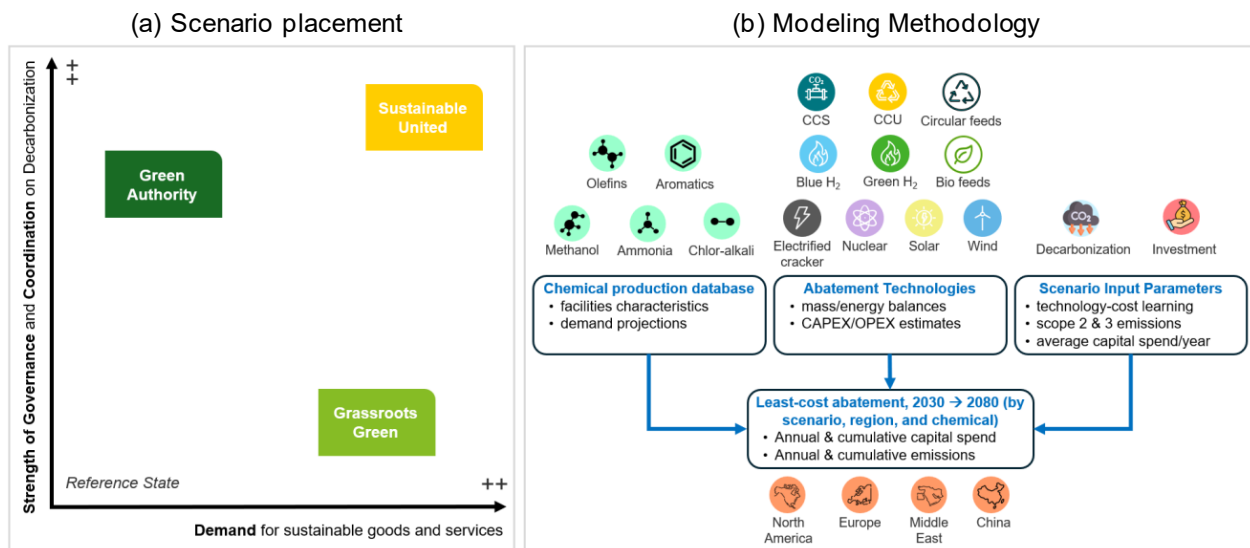

*Figure 1. Overview of modeling approach for this study. (a) Three scenarios placed notionally within the two-axis study framework: Sustainable United (SU), Green Authority (GA), and Grassroots Green (GG). (b) Summary of quantitative modeling methodology.*

The Grassroots Green (GG) scenario is characterized by fragmented governance around sustainability, but strong consumer-led demand for sustainable products. The latter drives only localized progress, which subdues innovation and limits technology transfer. Global tensions and national interests limit cooperation among nations and corporations, with "buy local" procurement policies driving domestic investment in manufacturing and innovation. Low rates of technology transfer and fragmented supply chains result in more significant cost premiums for advanced clean technologies than in the other two scenarios, but consumers, particularly in higher income brackets, show a willingness to pay premiums, driving some demand for sustainable goods.

**2.2. Quantitative modeling**

Figure 1b summarizes the quantitative modeling approach used to determine least-cost emissions abatement trajectories and associated capital investment requirements by region and chemical over a modeling time horizon to 2080 for each of the three study scenarios. Emissions estimated in this study are on a "well-to-gate" basis, including scope 1 (direct), 2 (from grid electricity generation), and upstream 3 (associated with fuel and feedstock supply).

**2.2.1. Chemical facilities database and projected production**

The modeling is anchored by a chemical industry database and chemicals demand projections provided on a proprietary basis for this study by Chemical Market Analytics / Oil Price Information Service (CMA/OPIS) operated by Dow Jones & Company [41].

The database includes 4,012 individual building block chemical production assets that were operating commercially in 2023. For each asset, the database includes company owner, geographical location, start-up year, chemical production process, production capacity and utilization, input feedstock type, per-unit production intensity of input feedstock, electricity, steam, fuel (excluding steam-raising fuel), and estimated process $CO_2$ emissions per unit of production (as distinct from $CO_2$ emissions from fuel combustion or steam raising). The production assets, using a wide variety of feedstocks and production processes (SI B10), are located in 2,676 distinct facilities (SI B1).

CMA/OPIS also provided regional demand projections to 2050 for each of the studied chemicals and corresponding regional capacity and production projections. These were extrapolated to 2080 by the authors. Estimates of new capacity (in aggregate for each region) needed to meet future demands were also projected. See SI A1 and B1. Projected production levels were assumed to be the same in each scenario described in Section 2.1.

### 2.2.2. Emission abatement technologies

Potential emission abatement levels at each facility were estimated considering different commercially-deployable approaches discussed in SI A2 and briefly summarized here. Process flow schematics for all abatement options are provided SI B22. Technology performance and cost characterizations used as modeling inputs are in SI B14 to B21.

For abatement via $CO_2$ capture and storage (CCS) amine-based post-combustion capture is assumed, with heat needed for solvent regeneration provided by natural gas cogeneration, the $CO_2$ emissions from which are also captured. A 95% capture rate is assumed for low-purity $CO_2$ streams. Process-derived $CO_2$ associated with ammonia and methanol production (as distinct from combustion emissions associated with heat supply) are assumed to be 100% captured by simple dehydration due to their high $CO_2$ purity. For underground storage, captured $CO_2$ is assumed to be compressed and moved by pipeline to suitable storage sites.

Electrification is considered as an abatement option for steam crackers but is not considered available for commercial deployment until 2040, given its current early stage of technological development [42].

For hydrogen ($H_2$) as an abatement option, blue $H_2$ (from autothermal reforming of natural gas with CCS) and green $H_2$ (produced by water electrolysis using electricity from a carbon-free source) are considered.

For electrified abatement options, we assume the production facility executes power purchase agreements (PPA) with a nuclear plant, wind farm, and/or solar farm located relatively nearby (e.g., in the same state or province). In estimating capital requirements for the electricity supply (discussed in Section 2.2.3), the capacity of the generating plant is set such that its annual generation matches the annual electricity demand of the abatement facility.

For methanol production, abatement by CCS and by $CO_2$ capture and utilization (CCU) are two of the options considered. For CCU, captured $CO_2$ is an input feedstock alongside green $H_2$. The $CO_2$ is assumed to be captured at a nearby industrial site that may or may not be in the chemical sector. The CCU option is considered only for future new-build units.

For ammonia, blue and green $H_2$ options are considered, but green $H_2$ is assumed to be only for new-build applications, not for retrofits.

Abatements by feedstock substitution are also considered. Circular feedstock (oil from pyrolysis of plastic waste) is assumed for ethylene production, and biogenic feedstocks are considered for ethylene production (by ethanol dehydration) and for methanol production (by gasification of lignocellulosic biomass to produce a synthesis gas feedstock). The bio-ethylene option is considered only for future new-build facilities. For the bio-syngas option, retrofits are assumed to be feasible at existing methanol facilities and at new builds, and 20% of syngas from incumbent feedstocks is assumed to be replaced by bio-syngas. The assumed level of circular feedstock use varies by scenario. In the SU, GA, and GG scenarios, these grow to reach 20%, 15%, and 10%, respectively, of all input feedstock for ethylene production in North America and Europe by 2050 and in China and Middle East by 2060.

Emissions associated with chlor-alkali production are primarily associated with the carbon footprint of the electricity supply to the facility, which is assumed to be from the grid. Assumed grid carbon footprint reductions over time are discussed in Section 2.2.5.

### 2.2.3. Abatement technology costs

In addition to emissions estimates, key facility-level analytical outputs include abatement capital investment requirements and levelized costs of abatement (LCOA), both of which depend on how much technology cost-learning has occurred for a particular abatement solution by the time a project is deployed. SI A3 details the methodology for LCOA estimation.

For each combination of chemical and abatement option, facility level calculations are carried out for a region assuming each facility adopts the same abatement option over time. For chemicals with more than one abatement technology option, each option is evaluated for each facility to determine which one offers the lowest LCOA. Industry-wide emissions estimates for each scenario adopt the least-cost abatement options among the alternatives evaluated.

In the case of abatement approaches involving procurement of clean electricity (nuclear, wind, or solar PV), the capital costs for electricity generation are assumed to be part of the required abatement capital. Reference capital cost estimates for abatement technologies deployed in North America are adjusted by location factors to estimate costs for other regions. See SI Table A2.

For abatement retrofits to existing production facilities, no assessment was made of the execution feasibility of retrofits (land availability, access, services capacity, etc.) The implicit assumption is that there would be sufficient physical space, and other requirements could be met for installing and operating abatement equipment. This may underestimate retrofit capital needs. It is additionally assumed that chemicals production is not disrupted during equipment installation, which may contribute to underestimating LCOA values.

Because the geographic locations of future new-build facilities needed in a region to meet projected production levels are not known, we assume new capacity is built in locations that provide the most favorable conditions for the abatement option deployed. For example, when CCS is deployed, the facility is assumed to be built in a location with the lowest $CO_2$ transport and storage costs.

For each abatement project (retrofit or new-build), the estimated CAPEX is assumed to be expended over the course of a project development time from conceptualization to initial commercial operation that varies between three and seven years, depending on the abatement technology, with annual capital outlays following logistics curves (SI B6). Facilities are assumed to operate at rated capacity immediately upon start of commercial operation and to continue to operate to 2080.

### 2.2.4. Abatement technology deployment rates

In each scenario, the first abatement project for each chemical in a region is assumed to begin operating in 2030. After the first abatement project in a region, capital cost reductions by cost-learning are reflected in subsequent projects, with learning rates that vary by technology and by scenario. For a given technology, cost-learning is assumed to be slower during an "early mover" phase of deployment [43] and to accelerate as commercial experience grows. Because of different levels of assumed information and technology sharing across scenarios the most-rapid learning occurs in the SU scenario and least-rapid in GG. SI A4 includes details of cost-learning assumptions, and SI B5 shows region-specific learning curves assumed for each technology in each scenario.

In tandem with cost learning, project deployment rates vary between the three scenarios. In the SU ("best-of-all-possible worlds") scenario, it is assumed that sufficient industry-wide investment capital for emissions abatement is available to retrofit all existing facilities in North America & Europe by 2050 and in China and the Middle East by 2060, consistent with announced commitments. (Any new capacity built to meet growing chemicals demand is also assumed to be built with abatement.) In the GG and GA scenarios, consistent with the scenario narratives, capital is deployed at notionally slower rates reflecting less favorable investment conditions, which limits the rate at which abatement projects are brought on line.

### 2.2.5. Scope 2 and upstream 3 emissions reductions

Cost learning and technology deployment trajectories are the factors that differ most significantly between scenarios, but additional quantitative differences include rates of reduction in emissions from grid electricity generation (Scope 2 emissions) and from extraction and delivery of feedstocks and fuels (upstream scope 3 emissions). Quantitative details are in SI A5.

## 3. Results

### 3.1. Least-cost abatement technology options

For each study region, LCOAs were calculated for each chemical facility with each abatement option suitable for that facility. Figure A2 in the SI identifies the least and second-least cost options for each chemical in each region, and SI B23 shows individual facility LCOAs over time for the least-cost options.

For olefins, blue $H_2$ provides the lowest cost abatement in North America. In the Middle East blue hydrogen is least costly initially, but the electrified cracker option using renewable electricity has the lowest abatement cost starting in 2040, the first year that cracker electrification is assumed to be commercially deployable in any region. In Europe, the relatively high cost of natural gas makes CCS less costly than blue $H_2$. In China, green $H_2$ produced using renewable electricity is the least cost option initially but gives way to cracker electrification from 2040. For production of propylene (not from steam cracking) and aromatics, methanol and ammonia, CCS is the least-cost option in all regions over the full-time frame of the analysis.

It is noted that neither circular nor biogenic feedstock substitution is a least-cost option in any region. Though not promising for abatement on a cost basis, feedstock substitution tends to be motivated by drivers other than abatement cost reduction, like societal preferences for reduced reliance on fossil fuels, minimized generation of plastic waste, or for recyclable or biodegradable products. Feedstock supply may limit substation strategies, as discussed in SI A7.

### 3.2. Case study results for North American steam crackers

To illustrate how capital availability constraints impact modeled emissions reduction trajectories, we present detailed results for steam cracker emissions abatement via blue $H_2$ in North America.

In the SU scenario, abating all 47 existing North American steam cracker facilities by 2050 requires annual capital expenditures as in Figure 2a, top. Expenditures in a given year are for multiple simultaneous projects in various stages of development and construction. The average investment from 2025 until all existing crackers are abated in 2050 is $3.2 billion per year (2024$) over and above annual business-as-usual capital spending. Facilities are deployed in order of lowest LCOA, taking into consideration that cost learning is related to the number of projects previously deployed. The facility with the lowest LCOA at any point in time tends to be the largest-capacity facility due to economies of scale.

Three abated facilities are assumed to be operating before 2035, requiring capital spending starting in 2024 and reaching 3 to 4 B$/y by the late 2020s. Levelized costs of abatement for these initial projects are around $200/t (Figure 2a, middle). Parallel development does not allow for cost learning between these projects so costs for each is "first-of-kind". But projects that come on line starting in 2035 benefit from cost learning, leading to LCOAs that decline over time until the mid-2040s, when most remaining unabated facilities are relatively small and/or distant from lower cost $CO_2$ storage resources, leading to increasing LCOAs. Capital spending in the early/mid 2040s is 4 to 7 B$/y (Figure 2a, top), before declining in the second half of that decade due to the lower absolute amount of capital required for abating smaller capacity facilities. Post-2050, average annual spending is modest compared with pre-2050 since investments are needed only to abate capacity expansions, which are assumed to be world-scale facilities built near low-cost $CO_2$ storage resources and benefit from cost learning, resulting in levelized costs around $150/t (Figure 2a, middle). Industry-wide emissions, including from capacity additions to meet demand growth, are reduced by about 90% by 2050 (Figure 2a, bottom). Even with 95% reduction of scope 1 emissions, residual upstream scope 3 emissions persist. The required CO2 storage rate in 2050 is 73 million t/y and grows to 100 Mt/y by 2080.

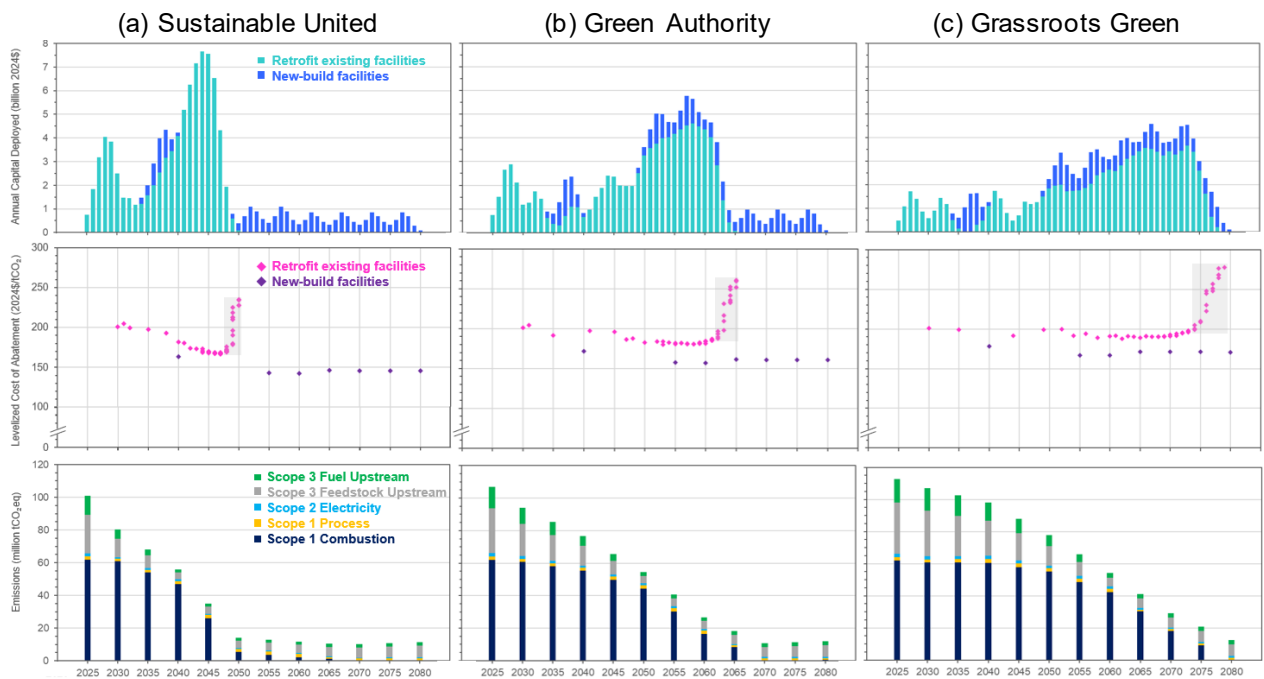

*Figure 2. Modeling results for abatement of steam cracker emissions in North America via blue H2 for (a) Sustainable United, (b) Green Authority, and (c) Grassroots Green. For each scenario, the panels from top to bottom are industry-wide annual abatement capital expenditure, facility-specific levelized cost of abatement (LCOA) for scope 1 emissions, and industry-wide emissions by scope. The LCOA values under the gray overlay rectangle are for facilities with small capacities or distant from low-cost CO2 storage. In practice, such facilities might be retired and collectively replaced by a smaller number of larger capacity facilities located near good storage resources.*

In the GA and GG scenarios we constrain average annual capital spending to $2.3 B$/year and 1.9 B$/y, respectfully, and cost learning is slowed because of the less collaborative environments than in SU. These factors combine to delay abatement of all existing crackers to

2063 and 2079, respectively (Figure 2b and Figure 2c, top), and push cumulative capital spending (2025 to 2080) to 92 B$ and 97 B$, respectively, from 86 B$ in SU for the same level of scope 1 emissions abated. With slower abatement trajectories, $CO_2$ storage requirements in 2050 in GA and GG are half and a third of those in SU, respectively, but reach 100 Mt/year by 2080 in both scenarios.

### 3.3. Global results

The same type of analysis for steam crackers in North America was applied across all chemicals and regions. Table 1 shows the annual average capital investment by chemical in each region and scenario. In the GG scenario for the Middle East and China, the expenditure rate is insufficient to abate all existing facilities within the modeled time horizon (2080).

In the case of North America, for which historical "business-as-usual" capital investment data are readily available, the annual investments to abate olefins and aromatics facilities can be compared with historical capital spending levels of the petrochemical manufacturing industry (North American Industry Classification System, NAICS code 325110). Average annual investment from 1990 to 2023 in this sector was 2.7 B$/y (SI B8). Average abatement capital expenditures for this sector in our scenarios range from 2.2 B$/y to 3.7 B$/y (Table 1), spending that would be above and beyond "business-as-usual" capital spending. For abatement of methanol and ammonia facilities in North America the average annual abatement investments are well below historical levels for the corresponding industry sectors, NAICS 325199 (Basic Organic Chemical Manufacturing not elsewhere classified, which includes methanol), 1.6 B$/y (1990 - 2023), and NAICS 325111 (nitrogenous fertilizer production, which includes ammonia), $3.6 B$/y (2005 - 2023), respectively (SI B9).

With the average annual capital expenditure rates in Table 1, the total cumulative capital deployed to abate all existing and new-build facilities across all regions from 2025 to 2080 is $1.1 to $1.2 trillion, for all three scenarios. (SI B7 disaggregates the cumulative total by region, by chemical, and by retrofit vs. new build.) China accounts for more than 50% of the spending, followed by North America, Europe, and the Middle East. Among the chemicals analyzed, olefins and aromatics account for more than 50% of the total in North America and Europe and for more than 60% in China and the Middle East.

Table 1. Average annual capital deployed in the Sustainable United, Green Authority, and Grassroots Green scenarios.

| Billion $/year (2024$) | North America | | | Europe | | | Middle East | | | China | | | Total | | |
|---|---|---|---|---|---|---|---|---|---|---|---|---|---|---|---|
| | SU | GA | GG | SU | GA | GG | SU | GA | GG | SU | GA | GG | SU | GA | GG |
| Olefins + aromatics | 3.7 | 2.7 | 2.2 | 2.7 | 2.0 | 1.6 | 1.2 | 0.9 | 0.9 | 12.1 | 8.5 | 6.3 | 19.8 | 14.1 | 11.0 |
| Steam crackers | 3.2 | 2.3 | 1.9 | 2.4 | 1.7 | 1.4 | 1.07 | 0.78 | 0.84 | 9.15 | 6.40 | 4.49 | 15.8 | 11.2 | 8.63 |
| On-purpose propylene | 0.36 | 0.26 | 0.20 | 0.26 | 0.19 | 0.15 | 0.12 | 0.10 | 0.07 | 2.60 | 1.87 | 1.56 | 3.34 | 2.42 | 1.98 |
| Aromatics | 0.15 | 0.11 | 0.09 | 0.08 | 0.06 | 0.05 | 0.05 | 0.05 | 0.03 | 0.34 | 0.25 | 0.21 | 0.62 | 0.47 | 0.38 |
| Methanol | 0.38 | 0.28 | 0.21 | 0.40 | 0.29 | 0.24 | 0.25 | 0.19 | 0.16 | 1.86 | 1.34 | 1.11 | 2.89 | 2.09 | 1.74 |
| Ammonia | 1.0 | 0.74 | 0.57 | 1.04 | 0.75 | 0.62 | 0.30 | 0.23 | 0.19 | 2.08 | 1.50 | 1.25 | 4.46 | 3.22 | 2.68 |
| Totals | 5.1 | 3.7 | 3.0 | 4.2 | 3.0 | 2.5 | 1.8 | 1.4 | 1.3 | 16.0 | 11.4 | 8.6 | 27.1 | 19.4 | 15.4 |
| Existing facilities abated | 2050 | 2063 | 2079 | 2050 | 2063 | 2080 | 2060 | 2080 | >2080 | 2060 | 2060 | >2080 | | | |

The cumulative capital investments appear consistent with a major earlier modeling study that estimated abatement capital investment requirements for a much larger scope of study (Christian Zibunas, 2022): full supply chain emissions abatement for 18 large-volume base chemicals plus 14 large-volume plastics and the treatment of corresponding plastic wastes, versus cradle-to-gate abatement for nine basic chemicals in our study (without downstream supply chain abatement). The scope of the earlier study was also global rather being limited to the four regions considered here, and it also assumed higher future percentage demand growth than assumed here.

Additionally, it assumed complete abatement of emissions instead of the maximum practicable level considered here.

Industry-wide emissions over time for the SU, GA, and GG scenarios are shown in Figure 3. For perspective, a reference case is also shown, with emissions intensities (scopes 1, 2, and upstream 3) "frozen" at the estimated 2023 levels, but with production growth the same as in the abatement scenarios. (See SI B1 for assumed frozen intensity values for each chemical in each region, and SI B2 for tabulated time-series of emissions by scenario, region, and chemical.) In the frozen-intensities case emissions collectively across the four regions grow by 70% from 2025, to 1.7 Gt/y by 2050, in line with projected annual production growth. Emissions exceed 2 Gt/y by 2080, when production is 2.5 times the 2025 level. Cumulative emissions from 2025 to 2080 are 95 Gt in the frozen-intensities case, compared to 27 Gt, 39 Gt, and 48 Gt in the SU, GA, and GG scenarios, respectively.

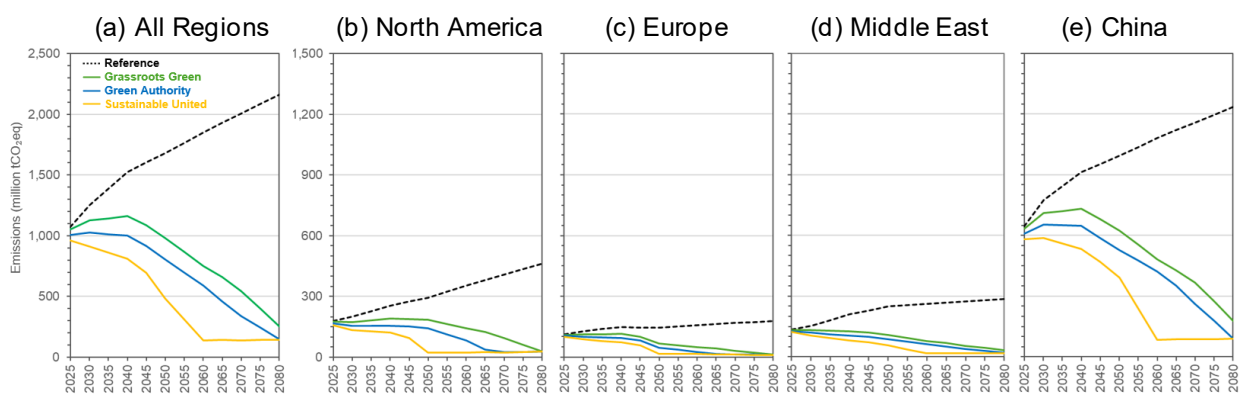

Figure 3. Total annual emissions (scope 1, 2, and upstream 3 for all included chemicals) by scenario. Reference case (REF) assumes emissions intensities frozen at 2023 levels. (Emissions diverge slightly between scenarios in 2025 because of scenario-based assumptions regarding reductions of scope 2 and scope 3 emissions relative to 2020 anchor values.)

In the SU scenario, global emissions decline 53% from 2025 to 2050, when all existing facilities in North America and Europe have been abated. Global emissions decline to 14% of 2025 emissions (140 MtCO2) by 2060, when all existing facilities in China and the Middle East have also been abated. Emissions rise slightly beyond 2060 due to added production units. These units are built with abatement but nevertheless have residual unabated emissions. In the GA scenario, with its slower abatement deployment rate, emissions due to production growth nearly balance abatements in Europe, North America, and Middle East until about 2040, while emissions rise in China during this period. Emissions then begin declining in all regions and reach full abatement by 2080. In the GG scenario, rising emissions are also observed until about 2040 and emissions decline more slowly after that than in the GA scenario. Emissions are not fully abated in China or the Middle East by 2080 in the GG scenario due to the slower pace of capital deployment.

Figure 4(a) shows total emissions disaggregated by scope. Scope 1 combustion and process emissions are an estimated 608 MtCO$_2$ today, or about two-thirds of the total emissions. In the SU scenario, by 2060 Scope 1 emissions are reduced by over 85%, but still account for about two-thirds of the total. Scope 2 emissions are nearly eliminated by 2060, since the grid is assumed to be largely decarbonized by then in all regions. On a global basis, scope 3 upstream emissions are assumed to be reduced by 95% from today's level by 2060 in the SU scenario, and these account for about one-third of the residual emissions of the industry in 2060. In the GA and GG scenarios scope 2 and scope 3 upstream emission reduction rates are less aggressive than in the SU scenario, and this compounds the slower rate of emissions reductions associated with reduced capital spending on abatement of scope 1 emissions.

Figure 4(b) disaggregates total emissions by chemical. Olefins account for the largest fraction of emissions across the time horizon analyzed, followed by ammonia, methanol, chlor-alkali, and aromatics. Notably, emissions for ammonia and methanol decline modestly in the SU scenario through the mid-2040s, while increasing initially in the GA and GG scenarios. These trends are related in large part to emissions contributions from China, where there is especially high projected production growth. China accounts for 40% to 60% of global production of both ammonia and methanol over the study time horizon. Production globally nearly doubles for methanol and more than doubles for ammonia from 2025 to 2060. Additional contributing factors include the target date for full abatement of existing facilities being 2060 rather than 2050 for China and the use of coal feedstocks in China, which results in higher emissions per unit of production from not-yet-abated facilities compared with natural gas-based production. The slower pace of abatement of existing higher carbon-intensity facilities and the added residual emissions from abated new facilities combine to keep emissions from methanol and ammonia production globally relatively flat until the late 2040s. In the GA and GG scenarios, the decline in emissions from the 2040s is slower relative to the pace in SU.

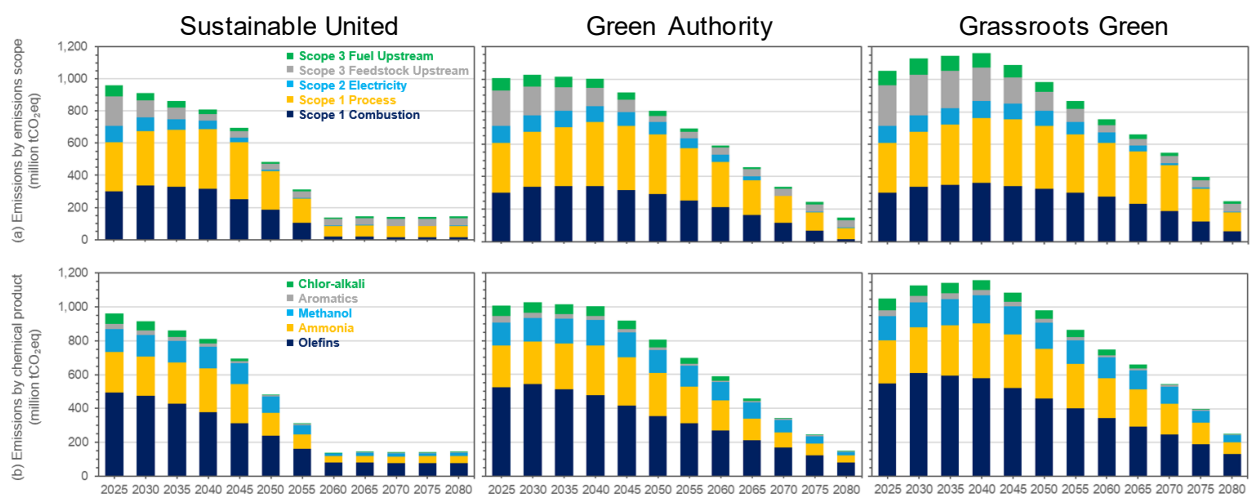

Figure 4. Emissions over time for three scenarios disaggregated by (a) emissions scope and (b) chemical product.

## 4. Conclusions and future work suggestions

Our over-arching results are sobering. In an ideal (Sustainable United) future, the abatement capital investment rate to deeply decarbonize the industry by mid-century (2050 in North America and Europe; 2060 in China and Middle East) is significantly greater than the chemical industry's historical "business-as-usual" (not for abatement) capital spending rate. In scenarios where governance and coordination and/or consumer demand and willingness to pay for sustainable production are less strong (Green Authority, Grassroots Green), lower capital deployment rates mean the industry would not fully decarbonize until well into the second half of this century, at best. An additional decarbonization challenge for any of the scenarios implied in the results, though not explicitly modeled, is the significant additional human capital that would be required – competent design engineers, specialized construction crews, trained operators – to develop, build, and operate abatement projects at the needed rates. Addressing such challenges will require supportive policies, for example public/private partnerships, regulated collaboration on abatement projects among competitors, trusted mechanisms for distributing the value of lower-carbon-intensity basic chemicals across the full chemicals value chains, and expanded workforce training programs, among others [44].

Our findings suggest several important directions for future work:

(1) We did not include in our analysis several regions with relatively small chemicals industries today, but with rapidly growing economies, including India and Southeast Asia. Understanding how chemicals industries might grow sustainably in such regions would be important.
(2) Potential future uses of hydrogen, methanol, or ammonia beyond current uses, e.g., for transportation or power generating fuel, have not been considered and so we may underestimate future demand and emissions associated with these chemicals. Future work would help clarify the implications of this omission for the decarbonization challenge.
(3) Our results draw attention to the important contribution of upstream scope 3 emissions to the chemical industry's total emissions today while assuming that pledges of the oil and gas producers to reduce these rapidly and dramatically will be met. Future work to identify actions needed to achieve verifiable reductions in upstream scope 3 emissions would be useful.
(4) We found CCS to be a key least-cost abatement option; region-specific analyses are needed of how $CO_2$ transport and storage infrastructure could be most effectively developed, considering local socio-techno-economic conditions, to meet chemical industry decarbonization needs.
(5) As low-carbon electricity supplies increase in the future, electrification options, especially electrified steam cracking but also green $H_2$ production, are prospectively key decarbonization options in some regions. The commercialization of these is nascent today and work to better understand critical paths to commercial application for the chemicals industry would be useful.
(6) The dominant role of China in production of chemicals and its higher carbon-intensity today are noteworthy, and further work to understand decarbonization pathways for China would be especially valuable.
(7) Our analysis highlights the fact that petrochemicals (olefins and aromatics) account for the majority of basic chemicals production on a volume basis globally. Many studies project demand for primary refinery products (gasoline and diesel) will decline with growing electrification of road transport as the world decarbonizes. The opportunities and challenges of this transition for production of olefins and aromatics needs to be better understood.

## 5. Acknowledgements

The work reported here originated during a broader study [44] conducted jointly with collaborators at Deloitte, LLC, including David Yankovitz, Robert Kumpf, Jessica Perkins, Haakan Jonsson, and others. We thank the Deloitte team for their collegial engagement throughout. We also thank many practitioners from the chemical industry and other experts who participated in one or more of the three day-long workshops organized jointly with Deloitte in the course of this work to solicit feedback on methods, assumptions, and interim results. Financial support for this work was provided by Deloitte and Princeton University's Andlinger Center for Energy and the Environment.

## 6. References

[1] American Chemistry Council, "2023 Guide to the Business of Chemistry," Washington, D.C., USA. 2023. https://www.americanchemistry.com/content/download/14468/file/2023-Guide-to-the-Business-of-Chemistry.pdf.

[2]     "GDP (Current US$) | Data," World Bank. Washington, DC, USA. 2025. Retrieved 15 August 2025. https://data.worldbank.org/indicator/NY.GDP.MKTP.CD

[3]     Levi, P. G., and Cullen, J. M., "Mapping Global Flows of Chemicals: From Fossil Fuel Feedstocks to Chemical Products," *Environmental Science and Technology*, Vol. 52, No. 4, 2018, pp. 1725–1734. https://doi.org/10.1021/acs.est.7b04573

[4]     OECD, "Global Plastics Outlook: Policy Scenario to 2060," OECD Publishing, Paris, https://doi.org/10.1787/aa1edf33-en., 2022.

[5]     Harrandt, J., Carus, M., and vom Berg, C., "Evaluation of Recent Reports on the Future of a Net-Zero Chemical Industry in 2050," nova-Institut GmbH, Hürth, Germany. 2024. https://renewable-carbon.eu/publications/product/evaluation-of-recent-reports-on-the-future-of-a-net-zero-chemical-industry-in-2050-pdf/. https://doi.org/10.52548/SXWV6083

[6]     Kätelhön, A., Meys, R., Deutz, S., Suh, S., and Bardow, A., "Climate Change Mitigation Potential of Carbon Capture and Utilization in the Chemical Industry," *Proceedings of the National Academy of Sciences of the United States of America*, Vol. 166, No. 23, 2019, pp. 11187–11194. https://doi.org/10.1073/pnas.1821029116

[7]     Ishii, N., Kanazawa, D., Stuchtey, M., Speelman, E., Herrmann, S., Kremer, A., Wagner, A., Leung, J., and Goult, P., "Planet Positive Chemicals: Pathways for the Chemical Industry to Enable a Sustainable Global Economy," The University of Tokyo, Japan. 2022.

[8]     Science Based Targets initiative (SBTi), "Science Based Target Setting in the Chemicals Sector: Status Report," London, UK. 2023.

[9]     European Chemical Industry Council (CEFIC), "The Antwerp Declaration for a European Industrial Deal," Brussels, Belgium. 2024. https://antwerp-declaration.eu/.

[10]    U.S. Department of Energy, "Transformative Pathways for U.S. Industry: Unlocking American Innovation," Washington, D.C., USA. 2025. Accessed on January 18, 2025. https://www.energy.gov/eere/iedo/articles/transformative-pathways-us-industry-unlocking-american-innovation.

[11]    UK Government, "Industrial Decarbonisation Strategy," ISBN 978-1-5286-2449-7. London, UK., 2021.

[12]    European Commission, "A European Chemicals Industry Action Plan," Strasbourg, France. 2025.

[13]    Bains, P., Psarras, P., and Wilcox, J., "CO2 Capture from the Industry Sector," *Progress in Energy and Combustion Science*. Volume 63, 146–172. https://doi.org/10.1016/j.pecs.2017.07.001

[14]    Energy Transition Commission, "Mission Possible: Reaching Net-Zero Carbon Emissions from Harder-to-Abate Sectors by Mid-Century," London, United Kingdom. 2018.

[15]    Hermanns, R., Oliveira De Lima, C., Vögler, O., Wilson, J., Zehnder, S., and Meys, R., "Pathways for the Global Chemical Industry to Climate Neutrality," International Council of Chemical Associations (ICCA) & Carbon Minds. London, UK. 2024.

[16] Saygin, D., and Gielen, D., "Zero-Emission Pathway for the Global Chemical and Petrochemical Sector," *Energies*, Vol. 14, No. 13, 2021. https://doi.org/10.3390/en14133772

[17] Fleiter, T., Herbst, A., Rehfeldt, M., and Arens, M., "Industrial Innovation: Pathways to Deep Decarbonisation of Industry. Part 2: Scenario Analysis and Pathways to Deep Decarbonisation," Fraunhofer Institute for Systems and Innovation Research (ISI). Karlsruhe, Germany. 2019.

[18] Cattry, A., Vuppanapalli, C., and Mallapragada, D. S., "Comparative Reactor, Process, Techno-Economic, and Life Cycle Emissions Assessment of Ethylene Production via Electrified and Thermal Steam Cracking." https://doi.org/10.26434/chemrxiv-2025-0nbs9

[19] Attwood, J., "2024 Levelized Cost of Net-Zero Materials: Net Zero Carriers A Persistent Green Premium," BloombergNEF, London, UK. 2024.

[20] Zibunas, C., Meys, R., Kätelhön, A., and Bardow, A., "Cost-Optimal Pathways towards Net-Zero Chemicals and Plastics Based on a Circular Carbon Economy," *Computers and Chemical Engineering*, Vol. 162, 2022. https://doi.org/10.1016/j.compchemeng.2022.107798

[21] World Economic Forum. Switzerland, 2021., "Towards Net-Zero Emissions Policy Priorities for Deployment of Low-Carbon Emitting Technologies in the Chemical Industry," 2021.

[22] Eryazici, I., Ramesh, N., and Villa, C., "Electrification of the Chemical Industry—Materials Innovations for a Lower Carbon Future," *MRS Bulletin*. 12. Volume 46, 1197–1204. https://doi.org/10.1557/s43577-021-00243-9

[23] Meys, R., Kätelhön, A., Bachmann, M., Winter, B., Zibunas, C., Suh, S., and Bardow, A., "Achieving Net-Zero Greenhouse Gas Emission Plastics by a Circular Carbon Economy."

[24] Vasta, K., "Chemical Recycling of Plastics (Part 1): Market Update," BloombergNEF, London, UK. 2023.

[25] Vasta, K., "Chemical Recycling of Plastics (Part 2): Technologies and Costs," BloombergNEF, London, UK. 2023.

[26] Jeswani, H., Krüger, C., Russ, M., Horlacher, M., Antony, F., Hann, S., and Azapagic, A., "Life Cycle Environmental Impacts of Chemical Recycling via Pyrolysis of Mixed Plastic Waste in Comparison with Mechanical Recycling and Energy Recovery," *Science of the Total Environment*, Vol. 769, 2021. https://doi.org/10.1016/j.scitotenv.2020.144483

[27] Ragaert, K., Delva, L., and Van Geem, K., "Mechanical and Chemical Recycling of Solid Plastic Waste," *Waste Management*. Volume 69, 24–58. https://doi.org/10.1016/j.wasman.2017.07.044

[28] SYSTEMIQ, "ReShaping Plastics: Pathways to A Circular Climate Neutral Plastics System in Europe," London, UK. 2022.

[29] Plastics Europe, "The Circular Economy for Plastics : A European Analysis," Brussels, Belgium. 2024.

[30] Plastics Europe, "The Plastics Transition: Our Industry's Roadmap for Plastics in Europe to Be Circular and Have Net-Zero Emissions by 2050," Brussels, Belgium. 2022.

[31] Energy Transitions Commission, "Bioresources within a Net-Zero Emissions Economy: Making a Sustainable Approach Possible," London, UK. 2021.

[32] International Energy Agency, "The Future of Petrochemicals: Towards More Sustainable Plastics and Fertilisers," Paris, France. 2018.

[33] Katelhön, A., Bardow, A., and Suh, S., "Stochastic Technology Choice Model for Consequential Life Cycle Assessment," *Environmental Science and Technology*, Vol. 50, No. 23, 2016, pp. 12575–12583. https://doi.org/10.1021/acs.est.6b04270

[34] Citation Teske, S., Niklas, S., Nagrath, K., Talwar, S., Atherton, A. N., and Guerrero Orbe, J., "Limit Global Warming to 1.5˚C -Sectoral Pathways & Key Performance Indicators," Institute for Sustainable Futures University of Technology Sydney. NSW, Australia. 2022.

[35] Teske, S., Niklas, S., Talwar, S., and Atherton, A., "1.5 °C Pathways for the Global Industry Classification (GICS) Sectors Chemicals, Aluminium, and Steel," *SN Applied Sciences*, Vol. 4, No. 4, 2022. https://doi.org/10.1007/s42452-022-05004-0

[36] Huyett, C., Peltier, M., Dass, A., Cangelose, B., and Payer, B., "Chemistry in Transition: Charting Solutions for a Low-Emissions Chemical Industry," Rocky Mountain Institute. Colorado, US. 2025. https://rmi.org/chemistry-in-transition-charting-solutions-for-a-low-emissions-chemical-industry/.

[37] Meng, F., Wagner, A., Kremer, A. B., Kanazawa, D., Leung, J. J., Goult, P., Guan, M., Herrmann, S., Speelman, E., Sauter, P., Lingeswaran, S., Stuchtey, M. M., Hansen, K., Masanet, E., Serrenho, A. C., Ishii, N., Kikuchi, Y., and Cullen, J. M., "Planet-Compatible Pathways for Transitioning the Chemical Industry," *Proceedings of the National Academy of Sciences of the United States of America*, Vol. 120, No. 8, 2023. https://doi.org/10.1073/pnas.2218294120

[38] King, C. W., "An Integrated Biophysical and Economic Modeling Framework for Long-Term Sustainability Analysis: The HARMONEY Model," *Ecological Economics*, Vol. 169, 2020. https://doi.org/10.1016/j.ecolecon.2019.106464

[39] Cullen, J. M., Meng, F., Allen, D., Christopher, P., Hamlin, C., Hamlin, P., Gao, Y., Jabarivelisdeh, B., Jahani, E., Jennings, E., Jin, E., Kimura, Y., King, C., Lupton, R., Masanet, E., Poole, A., Sadati, S., Saunders, C., Cabrera Serrenho, A., Vandermark, T., Zhang, W., and Bauer, F., "Carbon Clarity in the Petrochemical Supply Chain: A Critical Review," Carbon clarity in the global petrochemical supply chain (C-THRU). University of Cambridge, UK. 2022. c-thru.org/publications, 2022.

[40] Wack, P., "Scenarios: Uncharted Waters Ahead," Harvard Business Review 63, no. 5, pages 72–89. Massachusetts, USA. 1985. Retrieved 8 September 2025. https://hbr.org/1985/09/scenarios-uncharted-waters-ahead

[41] "World Analysis - Chemical Market Analytics By OPIS, a Dow Jones Company," Maryland, USA. 2025. Retrieved 26 August 2025. https://chemicalmarketanalytics.com/services/world-analysis/

[42] Bloomberg Finance L.P., "Industry Decarbonization Market Outlook 1H 2024: Policy Steps Up," New York City, USA. 2024.

[43] Greig, C., Garnett, A., Oesch, J., and Smart, S., "Guidelines for Scoping and Estimating Early Mover CCS Projects," The University of Queensland, Australia. 2014.

[44] Deloitte, "Pathways Toward Sustainability: A Roadmap for the Global Chemicals Industry," New York, USA. 2025. (in preparation).